\documentclass[conference]{IEEEtran}
\IEEEoverridecommandlockouts

\usepackage{cite}
\usepackage{amsmath,amssymb,amsfonts}
\usepackage{algorithmic}
\usepackage{graphicx}
\usepackage{textcomp}
\usepackage{xcolor}
\usepackage[nolist]{acronym}
\def\BibTeX{{\rm B\kern-.05em{\sc i\kern-.025em b}\kern-.08em
    T\kern-.1667em\lower.7ex\hbox{E}\kern-.125emX}}

\usepackage{mathrsfs}
\usepackage{algorithm}
\usepackage{array}
\usepackage[caption=false,font=normalsize,labelfont=sf,textfont=sf]{subfig}
\usepackage{textcomp}
\usepackage{stfloats}
\usepackage{url}
\usepackage{verbatim}
\usepackage{color,soul}
\usepackage{hyperref}
\usepackage{cite}
\usepackage[nolist]{acronym}
\usepackage{mathtools}
\usepackage{amssymb}
\usepackage{stfloats}
\usepackage{makecell}
\usepackage{nccmath}
\acrodefplural{HPA}[HPAs]{high power amplifiers}
\usepackage{yhmath}

\usepackage{placeins} 
\usepackage{optidef}  

\begin{acronym}
	\acro{SSR}{sum secrecy rate}
 \acro{mmWave}{milli-meter wave}
    \acro{MIMO}{multiple-input, multiple-output}
    \acro{AWGN}{additive white Gaussian noise}
    \acro{KPI}{Key Performance Indicator} 
    \acro{D2D}{device-to-device}
    \acro{ISAC}{integrated sensing and communication}
    \acro{DFRC}{dual-functional radar and communication}
    \acro{AoA}{angle of arrival}
    \acro{EIRP}{effective isotropic radiated power}
    \acro{AoAs}{angles of arrival}
    \acro{ToA}{time of arrival}
    \acro{EVM}{error vector magnitude}
    \acro{CPI}{coherent processing interval}
    \acro{eMBB}{enhanced mobile broadband}
    \acro{URLLC}{ultra-reliable low latency communications}
    \acro{mMTC}{massive machine type communications}
	\acro{QCQP}{quadratic constrained quadratic programming}
	\acro{BnB}{branch and bound}
	\acro{SER}{symbol error rate}
	\acro{LFM}{linear frequency modulation}
	\acro{BS}{base station}
	\acro{UE}{user equipment}
	\acro{DL}{downlink}
	\acro{UL}{uplink}
	\acro{CS}{compressed sensing}
	\acro{PR}{passive radar}
	\acro{LMS}{least mean squares}
	\acro{NLMS}{normalized least mean squares}
	\acro{RIS}{reconfigurable intelligent surface}
    \acro{UAV}{unmanned aerial vehicle}
    \acro{STARS}{simultaneously transmitting and reflecting surface}
	\acro{RISs}{reconfigurable intelligent surfaces}
	\acro{IRS}{intelligent reflecting surface}
	\acro{LISA}{large intelligent surface/antennas}
	\acro{OFDM}{orthogonal frequency-division multiplexing}
	\acro{EM}{electromagnetic}
	\acro{MSE}{mean squared error}
	\acro{XR}{extended reality}
	\acro{SNR}{signal-to-noise ratio}
	\acro{SRP}{successful recovery probability}
	\acro{CDF}{cumulative distribution function}
	\acro{ULA}{uniform linear array}
	\acro{RSS}{received signal strength}
	\acro{SNDR}{signal-to-noise-and-distortion ratio}
	\acro{DPI}{direct path interference}
	\acro{RPI}{reflected path interference}
	\acro{DE}{direct echo}
	\acro{RE}{RIS-resulted echo}
	\acro{PI}{path interference}
	\acro{ADC}{analog-to-digital converter}
	\acro{RF}{radio frequency}
	\acro{BCD}{block coordinate descent}
	\acro{BCCD}{block cyclic coordinate descent}
	\acro{RCG}{Riemannian conjugate gradient}
	\acro{SDP}{semi-definite program}
	\acro{EVD}{eigenvalue decomposition}
	\acro{RCS}{radar cross-section}
	\acro{STAR}{simultaneous transmitting and reflecting}
	\acro{LNA}{low noise amplifier}
    \acro{NOMA}{non-orthogonal multiple access}
    \acro{FM}{frequency modulation}
    \acro{SINR}{signal to interference plus noise ratio}
    \acro{V2X}{Vehicle-to-Everything}
    \acro{NLoS}{non-line-of-sight}
    \acro{LoS}{line-of-sight}
    \acro{AN}{artificial noise}
	\acro{SI}{self interference}
    \acro{MIMO}{multiple-input, multiple-output}
    \acro{AWGN}{additive white Gaussian noise}
    \acro{KPI}{key performance indicator}
    \acro{KPIs}{key performance indicators}  
    \acro{D2D}{device-to-device}
    \acro{ISAC}{integrated sensing and communications}
    \acro{DFRC}{dual-functional radar and communication}
    \acro{AoA}{angle of arrival}
    \acro{AoAs}{angles of arrival}
    \acro{MU-MIMO}{multi-user, multiple-input, multiple-output}
    \acro{ToA}{time of arrival}
    \acro{PCA}{principle component analysis}
    \acro{EVM}{error vector magnitude}
    \acro{CPI}{coherent processing interval}
    \acro{eMBB}{enhanced mobile broadband}
    \acro{URLLC}{ultra-reliable low latency communications}
    \acro{mMTC}{massive machine type communications}
	\acro{QCQP}{quadratic constrained quadratic programming}
	\acro{BnB}{branch and bound}
	\acro{SER}{symbol error rate}
	\acro{LFM}{linear frequency modulation}
	\acro{BS}{base station}
	\acro{SLNR}{signal-to-leakage-plus-noise ratio}
	\acro{UE}{user equipment}
	\acro{DL}{deep learning}
	\acro{CS}{compressed sensing}
	\acro{PR}{passive radar}
	\acro{LMS}{least mean squares}
	\acro{NLMS}{normalized least mean squares}
	\acro{ZF}{zero forcing}
	\acro{RIS}{reconfigurable intelligent surface}
	\acro{RISs}{reconfigurable intelligent surfaces}
	\acro{IRS}{intelligent reflecting surface}
	\acro{IRSs}{intelligent reflecting surfaces}
	\acro{NOMA}{non-orthogonal multiple access}
	\acro{TWRN}{two-way relay networks}
	\acro{LISA}{large intelligent surface/antennas}
	\acro{LISs}{large intelligent surfaces}
	\acro{OFDM}{orthogonal frequency-division multiplexing}
	\acro{EM}{electromagnetic}
	\acro{ISMR}{integrated sidelobe to mainlobe ratio}
	\acro{MSE}{mean squared error}
	\acro{SNR}{signal-to-noise ratio}
	\acro{IoT}{internet of things}
	\acro{SRP}{successful recovery probability}
	\acro{CDF}{cumulative distribution function}
	\acro{ULA}{uniform linear array}
	\acro{RSS}{received signal strength}
	\acro{SU}{single user}
	\acro{MU}{multi-user}
	\acro{CSI}{channel state information}
	\acro{OFDM}{orthogonal frequency-division multiplexing}
	\acro{DL}{downlink}
	\acro{i.i.d}{independently and identically distributed}
	\acro{UL}{uplink}
	\acro{DFT}{discrete Fourier transform}
    \acro{HPA}{high power amplifier}
    \acro{IBO}{input power back-off}
    \acro{MIMO}{multiple-input, multiple-output}
    \acro{PAPR}{peak-to-average power ratio}
    \acro{AWGN}{additive white Gaussian noise}
    \acro{KPI}{key performance indicator}
    \acro{FD}{full duplex}
    \acro{KPIs}{key performance indicators}  
    \acro{D2D}{device-to-device}
    \acro{ISAC}{integrated sensing and communications}
    \acro{DFRC}{dual-functional radar and communication}
    \acro{AoA}{angle of arrival}
    \acro{AoD}{angle of departure}
    \acro{ToA}{time of arrival}
    \acro{EVM}{error vector magnitude}
    \acro{CPI}{coherent processing interval}
    \acro{eMBB}{enhanced mobile broadband}
    \acro{URLLC}{ultra-reliable low latency communications}
    \acro{mMTC}{massive machine type communications}
	\acro{QCQP}{quadratic constrained quadratic programming}
	\acro{BnB}{branch and bound}
	\acro{SER}{symbol error rate}
	\acro{LFM}{linear frequency modulation}
	\acro{ADMM}{alternating direction method of multipliers}
	\acro{CCDF}{complementary cumulative distribution function}
	\acro{MISO}{multiple-input, single-output}
	\acro{CSI}{channel state information}
	\acro{LDPC}{low-density parity-check}
	\acro{BCC}{binary convolutional coding}
	\acro{IEEE}{Institute of Electrical and Electronics Engineers}
	\acro{ULA}{uniform linear antenna}
	\acro{B5G}{beyond 5G}
	\acro{MOOP}{multi-objective optimization problem}
	\acro{ML}{maximum likelihood}
	\acro{FML}{fused maximum likelihood}
	\acro{RF}{radio frequency}
	\acro{AM/AM}{amplitude modulation/amplitude modulation}
	\acro{AM/PM}{amplitude modulation/phase modulation}
	\acro{BS}{base station}
	\acro{MM}{majorization-minimization}
	\acro{LNCA}{$\ell$-norm cyclic algorithm}
	\acro{OCDM}{orthogonal chirp-division multiplexing}
	\acro{TR}{tone reservation}
	\acro{COCS}{consecutive ordered cyclic shifts}
	\acro{LS}{least squares}
	\acro{CVE}{coefficient of variation of envelopes}
	\acro{BSUM}{block successive upper-bound minimization}
	\acro{ICE}{iterative convex enhancement}
	\acro{SOA}{sequential optimization algorithm}
	\acro{BCD}{block coordinate descent}
	\acro{SINR}{signal to interference plus noise ratio}
	\acro{MICF}{modified iterative clipping and filtering}
	\acro{PSL}{peak side-lobe level}
	\acro{SDR}{semi-definite relaxation}
	\acro{KL}{Kullback-Leibler}
	\acro{ADSRP}{alternating direction sequential relaxation programming}
	\acro{MUSIC}{MUltiple SIgnal Classification}
	\acro{EVD}{eigenvalue decomposition}
	\acro{SVD}{singular value decomposition}
	\acro{LDGM}{low-density generator matrix}
	\acro{SAC}{sensing and communication}
    \acro{HD}{half duplex}
    \acro{DM}{directional modulation}
    \acro{CDA}{constellation decomposition array}
    \acro{IAB}{integrated access and back-haul}
    \acro{PLS}{physical layer security}
    \acro{THz}{terahertz}
    \acro{CU}{communication user}
    \acro{SID}{study item description}
    \acro{3GPP}{3rd Generation Partnership Project}
    \acro{CRB}{Cram\'er-Rao Bound}
    \acro{ISABC}{integrated sensing and backscatter communication}
    \acro{DoF}{degrees-of-freedom}
    \acro{IJTB}{iterative joint Taylor-BCCD}
\end{acronym}

\DeclareMathOperator{\ISMR}{ISMR}
\DeclareMathOperator{\trace}{tr}

\DeclareMathOperator{\SR}{SR}

\DeclareMathOperator{\dB}{dB}

\DeclareMathOperator{\dBm}{dBm}
\DeclareMathOperator{\dBi}{dBi}
\DeclareMathOperator{\meters}{m}
\DeclareMathOperator{\GHz}{GHz}
\DeclareMathOperator{\dBmpHz}{dBm/Hz}
\DeclareMathOperator{\bpspHz}{bps/Hz}

\usepackage{xcolor,cite,etoolbox}
\makeatletter 
\pretocmd\@bibitem{\color{black}\csname keycolor#1\endcsname}{}{\fail}
\newcommand\citecolor[1]{\@namedef{keycolor#1}{\color{blue}}}
\makeatother

\newcommand*\HSI{\pmb{H}_{\tt{SI}}}

\usepackage{titlesec}

\titlespacing{\subsection}{3pt}{*0.05}{*0.1}
\titlespacing{\subsubsection}{3pt}{*0.05}{*0.1}
\usepackage{fancyhdr}
\fancypagestyle{firststyle}{
	\fancyhf{}
	\fancyhead[L]{A. Boljevi\'c, M. Ying, A. Bazzi, T. S. Rappaport, and M. Chafii, "Sum Secrecy Rate Maximization for Secure ISAC," accepted in \textit{2026 IEEE 1st Annual Integrated Sensing and Communication Conference (ISAC)}, November 2026.}

}
\def\BibTeX{{\rm B\kern-.05em{\sc i\kern-.025em b}\kern-.08em
    T\kern-.1667em\lower.7ex\hbox{E}\kern-.125emX}}
\begin{document}

\title{Sum Secrecy Rate Maximization for Secure ISAC}

\markboth{Journal of \LaTeX\ Class Files,~Vol.~14, No.~8, August~2021}%
{Shell \MakeLowercase{\textit{et al.}}: A Sample Article Using IEEEtran.cls for IEEE Journals}
\author{\IEEEauthorblockN{Aleksandar Boljevi\'c\IEEEauthorrefmark{1}, 
Mingjun Ying\IEEEauthorrefmark{2}, 
Ahmad Bazzi\IEEEauthorrefmark{1}\IEEEauthorrefmark{2}, 
Theodore S. Rappaport\IEEEauthorrefmark{2},
Marwa Chafii\IEEEauthorrefmark{1}\IEEEauthorrefmark{2}}
\IEEEauthorblockA{\IEEEauthorrefmark{1}Engineering Division, New York University (NYU) Abu Dhabi, 129188, UAE.}
\IEEEauthorblockA{\IEEEauthorrefmark{2}NYU WIRELESS, NYU Tandon School of Engineering, Brooklyn, 11201, NY, USA.}
{\{ab10863, my2770, ahmad.bazzi, tsr, marwa.chafii\}@nyu.edu}}


\maketitle
\thispagestyle{firststyle}
\begin{abstract}
In an integrated sensing and communications (ISAC) system, targets may intercept information. We address this specific security issue in a full-duplex ISAC system with malicious eavesdroppers aiming to intercept uplink (UL) and downlink (DL) communication exchanges between the dual-functional radar and communication base station and legitimate communication users. We formulate an optimization framework to maximize the sum secrecy rate for both DL and UL, considering power budget constraints for sensing and communications. The optimization problem is non-convex, so we introduce an iterative joint Taylor-block cyclic coordinate descent (IJTB) method to approximate it as a convex problem. The IJTB method alternates between sub-problems: one for UL beamformers and another for UL power allocation, artificial noise covariance, and DL beamforming, using Taylor approximations to simplify optimization. Simulations demonstrate the effectiveness of our approach compared to standard benchmarks.
\end{abstract}

\section{Introduction}
\label{sec:introduction}
Beyond-5G/6G wireless systems are evolving to support precise sensing tasks, such as \textcolor{black}{imaging \cite{11173662}}, Wi-Fi sensing in smart environments, and vehicle-to-infrastructure, \textcolor{black}{with future systems expected to operate at terahertz frequencies to enable hyper-accurate sensing, imaging, and position location~\cite{8732419}}, crucial for applications in robotics and smart homes.
However, \ac{ISAC} within $6$G networks presents significant challenges, one of which is security. Since \ac{ISAC} waveforms must both transmit information and sense targets, security risks arise, i.e., while radar energy should focus on eavesdropper detection, communication rates must be sustained for legitimate users.
Advances in \ac{FD} communications can support \ac{ISAC} by enabling simultaneous \ac{DL} and \ac{UL} transmissions, under \ac{SI} mitigation. Recent work addresses \ac{FD} waveform design for \ac{ISAC} nodes to enhance both radar and communication. The consumed power for secure \ac{FD} \ac{ISAC} systems was explored in \cite{10373185}.\\
\indent \textcolor{black}{From a security perspective, the \ac{SSR} measures the total rate at which legitimate users can exchange information with the base station such that no portion of that information leaks to an eavesdropper. Intuitively, SSR is the gap between two rates: (i) the rate the legitimate users can reliably decode, and (ii) the rate at which an adversary could decode the same signal, then summed over all legitimate links. 
The higher the \ac{SSR}, the more secure the legitimate links are.
Indeed, the secrecy capacity was first explored in Wyner's seminal wiretap-channel model \cite{wyner1975wire}, where a legitimate link coexists with an eavesdropping link observing a degraded version of the transmission. 
The secrecy capacity is the largest rate at which information can flow to the legitimate receiver, while remaining asymptotically undecodable by the eavesdropper. 
\ac{SSR} is positive only when the legitimate link \ac{SINR} exceeds that of the eavesdropper's, and at \ac{mmWave} the dual-directional beam gain can provide that margin.
The \ac{SSR} considered in this paper is the natural multi-user extension of this single link quantity, aggregating per-link secrecy rates across all legitimate \ac{DL} and \ac{UL} communications under the threat of multiple eavesdroppers.
\ac{SSR} is crucial in low-power IoT sensors in smart factories reporting to a gateway, while a malicious drone hovers nearby, or vehicle-to-infrastructure exchanges where a passing eavesdropper tries to intercept information, but is simultaneously sensed by the base station, within the context of \ac{ISAC}.}\\
\indent This paper aims to design a secure \ac{FD} \ac{ISAC} system where a \ac{DFRC} \ac{BS} in \ac{FD} mode serves both \ac{DL} and \ac{UL} \acp{CU} while detecting eavesdroppers. Using an \ac{AN} signal under \ac{ISMR} constraints, our framework optimizes the \ac{FD} \ac{SSR}, balancing communication with radar sensing needs. 
\textcolor{black}{Given the non-convex nature of the problem, we propose the \ac{IJTB} method in order to maximize the overall sum secrecy rate of the legitimate communication links, while respecting limited power budgets and a sensing-quality requirement on the radar beampattern.}\\
\indent Our results demonstrate that the \ac{IJTB} method \textcolor{black}{\textbf{outperforms benchmarks}} for \ac{UL} \ac{SSR} regardless of eavesdropper proximity; \textbf{maintains higher rates} as \ac{UL} radial distance increases; \textbf{efficiently utilizes \ac{AN}} across distances, and \textbf{converges rapidly} within approximately $10$ iterations, showing competitive performance even under relaxed \ac{ISMR} constraints.
The extended version of this work is available in \cite{ourSubmittedPaper}.\\
 \indent \textbf{Notation:} Vectors are denoted by lowercase boldface (e.g., $\pmb{x}$) with $\ell_2$ norm $\Vert \pmb{x} \Vert$; $\pmb{0}$ is the all zero vector (or matrix when applicable). Matrices are uppercase boldface, with transpose and Hermitian transpose denoted $(.)^T$ and $(.)^H$. 
 Non-negative vectors are $\pmb{x} \succeq \pmb{0}$, positive semi-definite matrices as $\pmb{A} \succeq \pmb{0}$, and $\pmb{I}_N$ is the $N \times N$ identity. Expectation is denoted by $\mathbb{E}\lbrace \cdot \rbrace$.

\vspace{-0.25cm}
\section{System Model}
\label{sec:system-model}
\subsection{Full Duplex Radar and Communication Model}
\FloatBarrier
\begin{figure}[!t]
\centering
\includegraphics[width=3in]{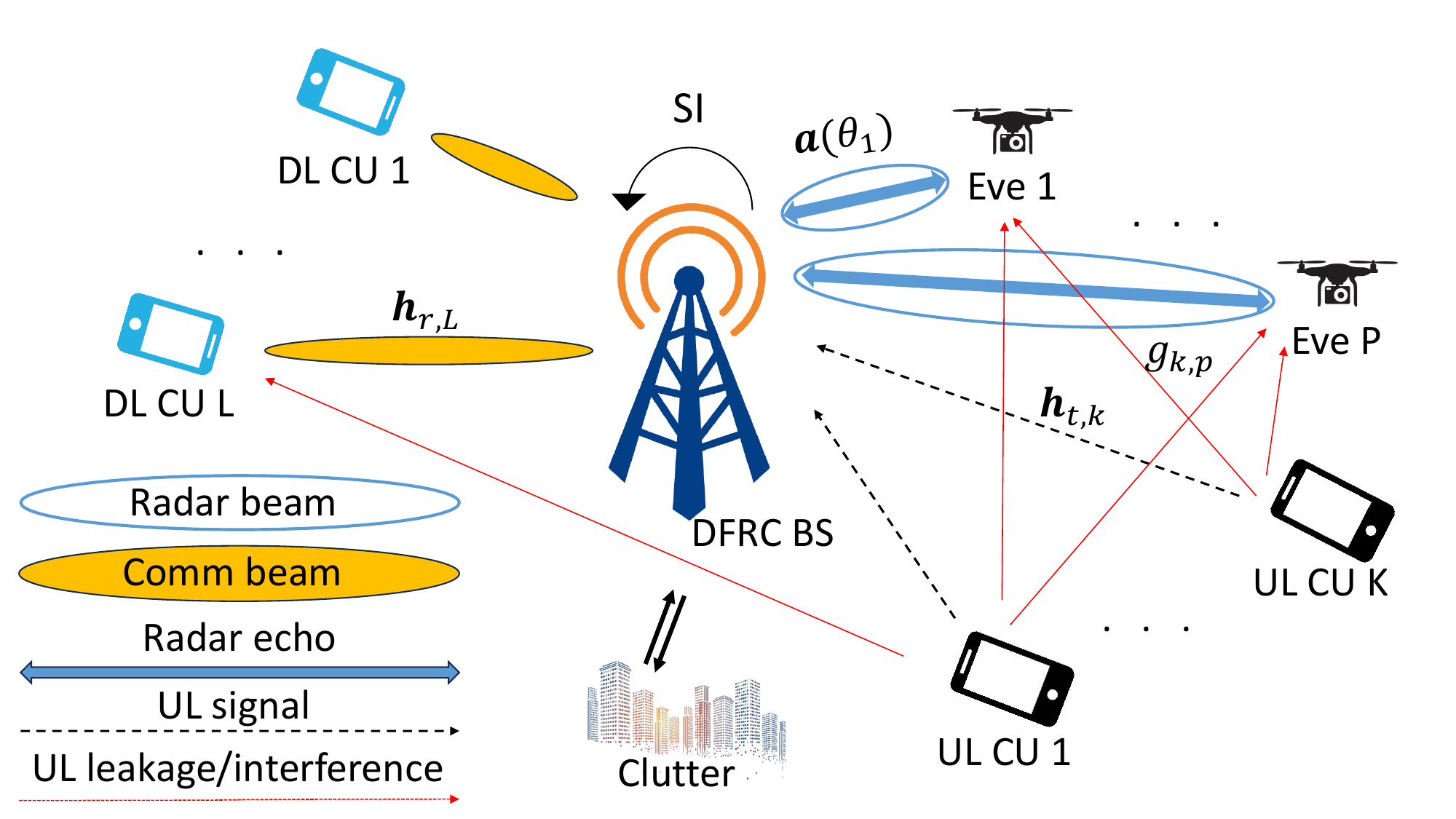}
\caption{The \ac{FD}-\ac{ISAC} system model: multiple malicious targets aiming to intercept communications between the \ac{DFRC} \ac{BS} and legitimate \ac{UL}/\ac{DL} \acp{CU}.}
\label{fig_1}
\end{figure}

\label{sec:FD-DFRC-model}
We consider an \ac{FD}-\ac{ISAC} setup, where a \ac{DFRC} \ac{BS} serves $K$ \ac{UL} and $L$ \ac{DL} legitimate \acp{CU}, with $P$ potential eavesdroppers attempting to intercept signals, as shown in Fig.\ref{fig_1}. The \ac{DL} signal is an \ac{ISAC} signal that also enables the \ac{DFRC} \ac{BS} to gather physical sensing data on eavesdroppers through directed beams. The \ac{DFRC} \ac{BS} is equipped with $N_T$ transmitting and $N_R$ receiving antennas in a monostatic radar setting. The transmitted secure \ac{ISAC} baseband signal vector $\pmb{x} = \pmb{Vs} + \pmb{w}$ consists of: \textit{(i)} a beamforming matrix $\pmb{V} \in \mathbb{C}^{N_T \times L}$, where each column directs signals to a \ac{DL} \ac{CU}, and \textit{(ii)} the data symbol vector $\pmb{s}$, containing independent symbols for each \ac{DL} \ac{CU} with unit variance. To enhance secrecy, an \ac{AN} vector $\pmb{w} \sim \mathcal{N}(\pmb{0},\pmb{W})$ with covariance $\pmb{W} \succeq \pmb{0}$ is added, where $\trace(\pmb{W})$ indicates the total power allocated to \ac{AN}. In addition, $\pmb{V}$ is optimized for joint DFRC beamforming, supporting both secure communication and radar sensing functions. Thus, while $\pmb{x}$ is primarily intended for \ac{DL} \acp{CU}, it also serves a radar function via $\pmb{V}$ and $\pmb{W}$, reinforcing security and sensing capabilities.
To this end, the single-antenna DL \acp{CU} receive the following signal
\useshortskip
\begin{equation}
 	\label{eq:DL-signal-2}
 	\small
		{y}^{\mathrm{DL}}_\ell
	=
	\pmb{h}_{r,\ell}^H
	\pmb{v}_\ell s_\ell
	+
	\sum\nolimits_{k=1}^K
	q_{k,\ell}v_k
	+
	\sum\nolimits_{\substack{\ell^{'} \neq \ell}}^L
	\pmb{h}_{r,\ell}^H
	\pmb{v}_{\ell^{'}} s_{\ell^{'}}
	+
	\pmb{h}_{r,\ell}^H
	\pmb{w}
	+
	n_\ell,
\end{equation}
where all terms are defined except for $n_\ell$, which follows $n_\ell \sim \mathcal{N}(0,\sigma_{\ell}^2)$. 
Moreover, $\pmb{h}_{r,\ell}$ denotes the \ac{mmWave} channel between the $\ell^{th}$ \ac{DL} \ac{CU} and the \textcolor{black}{DFRC BS \cite{rappaport2015millimeter}.}
The steering vector departing from \ac{AoD} $\varphi$ is denoted by $\pmb{a}_{N_T}(\varphi) \in \mathbb{C}^{N_T \times 1}$.
The channel coefficient between the $k^{th}$ \ac{UL} \ac{CU} and the $\ell^{th}$ \ac{DL} user is $q_{k,\ell}$, which captures the \ac{UL}-to-\ac{DL} interference. 
The $k^{th}$ \ac{UL} symbol is denoted as $v_k \in \mathbb{C}$, with power level $\mathbb{E}(\vert v_k \vert^2) = p_k$. 
Due to the presence of clutter, a portion of the \ac{DL} signal scatters from the clutter towards the \ac{DL} \acp{CU}, which is contained within the \ac{DL} channels $\pmb{h}_{r,\ell}$.
To this end, the $p^{th}$ eavesdropper reads
$	y_p^{\mathrm{Eve}}
	=
	\sum\nolimits_{k=1}^K
	g_{k,p}v_k
	+
	\alpha_p
	\pmb{a}_{N_T}^H(\theta_p)
	\pmb{x}
	+
	z_p$,  
where $g_{k,p}$ is the complex channel coefficient between the $k^{th}$ \ac{UL} \ac{CU} and the $p^{th}$ eavesdropper. The path-loss complex coefficient between the DFRC BS and the $p^{th}$ eavesdropper is denoted by $\alpha_p$.
The steering vector arriving at \ac{AoA} $\theta$ is denoted by $\pmb{a}_{N_R}(\theta) \in \mathbb{C}^{N_R \times 1}$.
The angles of arrival (AoAs) for the eavesdroppers are denoted as $\theta_p$.
The AWGN at the $p^{th}$ eavesdropper is $z_p$, modeled as zero-mean with variance $\sigma_p^2$, i.e., $z_p \sim \mathcal{N}(0,\sigma_p^2)$.
The \ac{DL} channel between \textcolor{black}{the} $\ell^{th}$ eavesdropper and the \ac{DFRC} \ac{BS} is denoted as $ \pmb{h}_{r,\ell} $, while the $k^{th}$ \ac{UL} \ac{mmWave} channel \textcolor{black}{\cite{rappaport2015millimeter}} is $ \pmb{h}_{t,k} $.
The eavesdropper, being the target, is assumed to have a strong \ac{LoS}\footnote{Dual-directional gain can be used to close the \ac{mmWave} link budget \cite{6387266}.} channel between itself and the \ac{DFRC} \ac{BS}, i.e. the channel is $ \pmb{a}(\theta_p) $, where $\pmb{a}(\theta)$ is the steering vector and \textcolor{black}{depends on the array} configuration\textcolor{black}{\cite{rappaport2015millimeter}}.
Although in \ac{ISAC} the \ac{AoA} must be sensed, and accuracy depends on array calibration \cite{10918620,ying2026site}, this work assumes perfect knowledge to quantify the optimistic performance the framework can give.
Additionally, the \ac{DFRC} \ac{BS} aims to direct sensing power towards $P$ eavesdroppers so the  backscattered echo returns experience good sensing \ac{SNR}.
The signal at the DFRC BS is
\begin{equation}
	\label{eq:BS-signal}
	 	\small
	\pmb{y}^{\mathrm{DFRC}}
	=
	\sum\nolimits_{k=1}^K \pmb{h}_{t,k} v_k
	+
	\pmb{y}_{\rm{e}+\rm{c}}
	+
	\sqrt{\beta}\HSI \pmb{x}
	+
	\pmb{n}.
\end{equation}
The first part in \eqref{eq:BS-signal} is the signal due to the $K$ \ac{UL} \acp{CU}.
The second part contains echo-plus-clutter returns, i.e. $\pmb{y}_{\rm{e}+\rm{c}} = 
	\sum\nolimits_{p=1}^P
	\gamma_p \pmb{a}_{N_R}(\theta_p)\pmb{a}_{N_T}^H(\theta_p)\pmb{x} + \pmb{c}$, 
where $\gamma_p$ is the complex amplitude due to the $p^{th}$ eavesdropper and $\pmb{c} \sim \mathcal{N}(\pmb{0},\pmb{R}_c)$ is the clutter. 
We assume the clutter covariance matrix $\pmb{R}_c \in \mathbb{C}^{N_R \times N_R}$, capturing all environmental clutter effects, to be constant during the coherent processing interval, and we further assume the eavesdropper \acp{AoA} are known.
The third part represents \ac{SI}, where $ \beta $ represents the residual \ac{SI} power and $\HSI$ represents the small-scale fading effects of \ac{SI}.
The last part is \ac{AWGN}, which follows $\pmb{n} \sim \mathcal{N}(\pmb{0},\sigma_n^2 \pmb{I}_{N_R})$.

\subsection{Key Performance Indicators (KPIs)}
\subsubsection{Secrecy rates}
We define the \ac{SINR} of the $\ell^{th}$ \ac{DL} \ac{CU}. Based on \eqref{eq:DL-signal-2}, assuming independent symbols and defining $\pmb{V}_{\ell} = \pmb{v}_{\ell} \pmb{v}^H_{\ell}$, the \ac{SINR} for the $\ell^{th}$ desired symbol, $s_{\ell}$, can be verified to be

\useshortskip
\begin{equation*}
	\label{eq:SINR_DL}
		  	\small
	\Gamma_{\ell}^{\mathrm{DL}}
	= 
	\frac{ \pmb{h}_{r,\ell}^H\pmb{V}_\ell\pmb{h}_{r,\ell} } 
	     {\sum\nolimits_{\substack{k = 1  }}^K p_k \vert q_{k,\ell} \vert^2 + \pmb{h}_{r,\ell}^H \left( \sum\nolimits_{\substack{\ell^{'} \neq \ell}}^L\pmb{V}_{\ell^{'}} + \pmb{W} \right)\pmb{h}_{r,\ell}   +\sigma_{\ell}^2}, 
\end{equation*}
Next, the achievable transmission rate of the $\ell^{th}$ \ac{DL} \ac{CU} is $
	R_{\ell}^{\mathrm{DL}} = \log_2( 1 + \Gamma_{\ell}^{\mathrm{DL}} ), \quad \forall \ell = 1 \ldots L. $
To evaluate \ac{UL} communication performance at the DFRC BS, we also define an \ac{UL} \ac{SINR}. Assuming the \ac{BS} applies $K$ beamforming vectors $\{\pmb{u}_k\}_{k=1}^{K}$ onto the receive signal in \eqref{eq:BS-signal}, we define a receive \ac{SINR} quantifying the performance between the $k^{th}$ \ac{UL} \ac{CU} and the \ac{BS} intending to detect $v_k$:

\useshortskip
\begin{equation}
	\label{eq:SINR_UL1}
	\small
	\Gamma_{k}^{\mathrm{UL}}
	=
    [p_k\pmb{u}_k^H \pmb{h}_{t,k}\pmb{h}^H_{t,k} \pmb{u}_k][\pmb{u}_k^H\pmb{\Phi}_k(\pmb{Q},\pmb{p})\pmb{u}_k]^{-1},
\end{equation}
where  
\useshortskip
\begin{equation}
\label{eq:phi_k_eqn}
\small
\pmb{\Phi}_k(\pmb{Q},\pmb{p}) = \sum\nolimits_{\substack{ k' \neq k }}^K  p_{k'}\pmb{h}_{t,k'}\pmb{h}^H_{t,k'}   +
	 \pmb{C}\pmb{Q}\pmb{C}^H + 
	 \pmb{R}_c +
	\sigma_n^2 \pmb{I},
\end{equation}
$\small\pmb{C} = \sum\nolimits_{p=1}^P \gamma_p \pmb{a}_{N_R}(\theta_p)\pmb{a}_{N_T}^H(\theta_p) + \sqrt{\beta}\HSI $; $\small\pmb{Q} = \pmb{V}\pmb{V}^H + \pmb{W}$,
Similarly, we can define the achievable rate of transmission of the $k^{th}$ \ac{UL} \ac{CU} as $\small R_{k}^{\mathrm{UL}} = \log_2( 1 + \Gamma_{k}^{\mathrm{UL}} ), \quad \forall k = 1 \ldots K$.
Prior to expressing the \ac{UL} and \ac{DL} \textit{secrecy rates}, we define a receive \ac{SINR} between the $\small p^{th}$ eavesdropper and the $\small k^{th}$ \ac{UL} \ac{CU}, who aims at detecting $v_k$. After some operations, it can be proven that the corresponding \ac{SINR} reads
\begin{equation*}
\small
	\Gamma_{p,k}^{\mathrm{Eve}}
	=
	\frac
	{ p_k \vert g_{k,p} \vert^2  }
	{\sum\nolimits_{\substack{ k' \neq k }}^K p_{k'} \vert g_{k',p} \vert^2 + \vert \alpha_p \vert^2 \pmb{a}_{N_T}^H(\theta_p)\pmb{Q}\pmb{a}_{N_T}(\theta_p) + \sigma_p^2}.
\end{equation*}
Similarly, we can express the receive \ac{SINR} between the $p^{th}$ eavesdropper and the \ac{DFRC} \ac{BS},
\useshortskip
\begin{equation*}
\small
	\Gamma_{p,\mathrm{DFRC}}^{\mathrm{Eve}}
	=
	\frac
	{\vert \alpha_p \vert^2 \pmb{a}_{N_T}^H(\theta_p)\pmb{VV}^H\pmb{a}_{N_T}(\theta_p)}
	{\sum\nolimits_{\substack{k = 1  }}^K p_k \vert g_{k,p} \vert^2 + 
	\vert \alpha_p \vert^2 \pmb{a}_{N_T}^H(\theta_p)\pmb{W}\pmb{a}_{N_T}(\theta_p)
	 + \sigma_p^2 }.
\end{equation*}
Having their \ac{SINR} expressions per eavesdropper defined, we will now define the \ac{DL} and \ac{UL} \acp{SSR} following \cite{7086319} as
\useshortskip
\begin{equation}
\label{eq:sum-secrecy-rate-DL}
\small
	\SR_{\mathrm{DL}}
	=
	\left[
	\sum\nolimits_\ell
	\log_2(1+\Gamma_{\ell}^{\mathrm{DL}})
	-
	\sum\nolimits_p
	\log_2(1+\Gamma_{p,\mathrm{DFRC}}^{\mathrm{Eve}})\right]^+.
\end{equation}
\useshortskip
\begin{equation}
\label{eq:sum-secrecy-rate-UL}
\small
\SR_{\mathrm{UL}}
	=\left[\sum\nolimits_{k}
	\log_2(1+\Gamma_{k}^{\mathrm{UL}})
	-
	\sum\nolimits_{p,k}
	\log_2(1+\Gamma_{p,k}^{\mathrm{Eve}})\right]^+.
\end{equation}
\subsubsection{Sensing metrics}
The \ac{DFRC} \ac{BS} uses its transmit vectors, $\pmb{x}$, to direct radar beams at $P$ eavesdroppers simultaneously, which enables the \ac{DFRC} \ac{BS} to gather environmental data from backscattered echoes and analyze each eavesdropper's physical attributes. We use the \ac{ISMR} KPI to evaluate radar performance, which quantifies the energy in the sidelobes relative to the mainlobe
\useshortskip
\begin{equation}
\label{eq:ISMR_def}
\small
	\ISMR \triangleq \frac
	{\int\nolimits_{\pmb{\Theta}_s}\pmb{a}_{N_T}^H(\theta) \pmb{R}_{\pmb{xx}} \pmb{a}_{N_T}(\theta) \ d\theta }
	{\int\nolimits_{\pmb{\Theta}_m}\pmb{a}_{N_T}^H(\theta) \pmb{R}_{\pmb{xx}} \pmb{a}_{N_T}(\theta) \ d\theta }
=
\frac
	{ \trace\big( \pmb{R}_{\pmb{xx}} \pmb{A}_s \big) }
	{\trace\big( \pmb{R}_{\pmb{xx}} \pmb{A}_m \big)},
\end{equation}
\vspace{-0.1cm}
where $\pmb{\Theta}_s$ and $\pmb{\Theta}_m$ are sets containing the directions toward the sidelobe and mainlobe components, respectively. Moreover, $\pmb{R}_{\pmb{xx}}$ represents the transmit covariance matrix, i.e, $\pmb{R}_{\pmb{xx}}
	=
	\pmb{V}\pmb{V}^H
	+
	\pmb{W}$.
The \ac{ISMR} could be reformulated after integrating over directions as per \eqref{eq:ISMR_def}, namely $\pmb{A}_{s /m} = \int\nolimits_{\pmb{\Theta}_{s/m}} \pmb{a}_{N_T}(\theta)\pmb{a}_{N_T}^H(\theta) \ d\theta  $. In this paper, $P$ mainlobes are centered around $\theta_1 \ldots \theta_P$, respectively. The mainlobe is given as
	$\Theta_m = \bigcup\nolimits_{p=1}^{P} \big\lbrace [\theta_p - \theta_p^{\mathrm{low}}; \theta_p - \theta_p^{\mathrm{high}} ] \big\rbrace$, 
where $\theta_p^{\mathrm{low}}$, $\theta_p^{\mathrm{high}}$ define the lower/upper angular regions of the $p^{th}$ mainlobe.
\section{Secure FD-ISAC SSR Optimization}
\label{sec:optimization-framework}
\subsection{FD-ISAC SSR Optimization Problem Formulation}
Given the power budget constraints for \ac{DL} beamforming, \ac{UL} power allocations, and \ac{AN} generation, we aim at maximizing the total \ac{SSR}s of both \ac{DL} and \ac{UL} \ac{CU}s. Therefore, we formulate the following optimization problem
\useshortskip

\makeatletter
\newcommand{\inlinenum}[1]{%
  \refstepcounter{equation}%
  \text{ (\theequation)}%
  \begingroup\@bsphack
    \protected@write\@auxout{}{%
      \string\newlabel{#1}{{\theequation}{\thepage}{}{}{}}}%
  \@esphack\endgroup
}
\makeatother
  \setlength{\jot}{2pt}                       
\begin{subequations}\label{eq:P1}
\begin{align}
\small
(\mathcal{P}_{1}):\quad
  & \underset{\pmb{p},\pmb{W},\pmb{V},\pmb{U}}{\operatorname{maximize}}
    \quad \SR_{\mathrm{DL}}+\SR_{\mathrm{UL}} \label{eq:P10}\\
\operatorname{s.t.}\quad
  & \trace(\pmb{VV}^H)\leq P_{\rm{DL}}^V \inlinenum{eq:P101},\ \
    \trace(\pmb{W})\leq P_{\rm{DL}}^W \inlinenum{eq:P102} \notag\\
  & \textstyle\sum_{k=1}^{K} p_k \leq P_{\rm{UL}} \inlinenum{eq:P103},\ \
    \ISMR \leq \ISMR_{\rm{max}} \inlinenum{eq:P104} \notag\\
  & \pmb{W}\succeq\pmb{0} \inlinenum{eq:P105},\ \
    \pmb{p}\succeq\pmb{0}. \inlinenum{eq:P106} \notag
\end{align}
\end{subequations}
\useshortskip
Constraints \eqref{eq:P101} to \eqref{eq:P103} impose power budgets on \ac{DL} beamforming ($P_{\rm{DL}}^V $), \ac{AN} generation ($P_{\rm{DL}}^W$), and \ac{UL} communication ($P_{\rm{UL}}$), respectively. 
Moreover, constraint \eqref{eq:P104} limits the maximum \ac{ISMR} to $\ISMR_{\rm{max}}$, while \eqref{eq:P105} ensures the covariance matrix $\pmb{W}$ is positive semi-definite and that the \ac{UL} power allocation vector has non-negative entries.
The problem is non-convex partly due to the $L$ rank-$1$ matrices $\sum\nolimits_\ell \pmb{V}_\ell = \pmb{VV}^H$, which can be relaxed as follows
\useshortskip
\begin{subequations}\label{eq:P11}
\begin{align}
\small
(\mathcal{P}_{1.1}):\quad
  & \underset{\pmb{p},\pmb{W},\{\pmb{V}_\ell\}_{\ell=1}^{L},\pmb{U}}{\operatorname{maximize}}
    \quad \SR_{\mathrm{DL}}+\SR_{\mathrm{UL}} \label{eq:P110}\\
\operatorname{s.t.}\quad
  & \textstyle\sum_{\ell=1}^{L}\trace(\pmb{V}_\ell)\leq P_{\rm{DL}}^V \inlinenum{eq:P111},\ \
    \pmb{V}_\ell\succeq\pmb{0},\ \forall\ell \inlinenum{eq:P112} \notag\\
  & \eqref{eq:P102},\ \eqref{eq:P103},\ \eqref{eq:P104},\ \eqref{eq:P105}. \notag
\end{align}
\end{subequations}
where the rank-one solutions are obtained using the strongest eigenvector of the solution corresponding to $\pmb{V}_\ell$.
However, $(\mathcal{P}_{1.1})$ is still non-convex due to the \ac{SSR} cost function, so we introduce a method to efficiently solve $(\mathcal{P}_{1.1})$.

\subsection{Iterative Joint Taylor-BCCD type method}
\label{sec:algorithmic-derivation}
In this section, we propose IJTB, i.e. iterative joint Taylor-BCCD, as a potential algorithm that efficiently solves the \ac{FD}-\ac{ISAC} \ac{SSR} maximization problem formulated via $(\mathcal{P}_1)$ through its relaxed version $(\mathcal{P}_{1.1})$.
To design IJTB, we must first convexify the problem $(\mathcal{P}_{1.1})$ so as to successfully iterate between the \ac{UL}/\ac{DL} beamformers, \ac{AN} statistics, and \ac{UL} power vector. 
For the sake of compact representation, we define the interference received by the $\ell ^{\rm{th}}$ 
\ac{DL} \ac{CU}, and the interference at the $p^{\rm{th}}$ eavesdropper as
\begin{align*}
\small
	&{\rm{I}}_{\ell}^{\mathrm{DL}} = 
	\sum\nolimits_{\substack{k }} p_k \vert q_{k,\ell} \vert^2 +\sum\nolimits_{\substack{ \ell' \neq \ell}} \trace \left( \pmb{V}_{\ell'}\pmb{H}_{r,\ell} \right) +\trace \left(\pmb{W}\pmb{H}_{r,\ell} \right), \\
&{\rm{I}}_{p, \mathrm{DFRC}}^{\mathrm{Eve}} = 
	\sum\nolimits_{\substack{k }} p_k \vert g_{k,p} \vert^2 + 
	\vert \alpha_p \vert^2 \trace \left(\pmb{W}\pmb{A}(\theta_p)\right),
\end{align*}
respectively.
The corresponding useful signal components of the $\ell ^{\rm{th}}$ DL \ac{CU} and the $p^{\rm{th}}$ eavesdropper are then ${\rm{S}}_{\ell}^{\mathrm{DL}} = \trace(\pmb{V}_{\ell}\pmb{H}_{r,\ell})$ and ${\rm{S}}_{p, \mathrm{DFRC}}^{\mathrm{Eve}} = \sum_\ell \trace \left(\vert \alpha_p \vert^2 \pmb{V}_\ell\pmb{A}(\theta_p)\right)$, respectively.
Therefore, the \ac{DL} \ac{SSR} can now be arranged as
\useshortskip
\begin{equation*}
\small
	\label{eq:SR_DL_rearranged}
 \begin{aligned}
 &\SR_{\mathrm{DL}} = 
 \sum_{\ell}\log _2\left({\rm{S}}_{\ell}^{\mathrm{DL}} + {\rm{I}}_{\ell}^{\mathrm{DL}} + \sigma_{\ell}^2 \right) + \sum_{p} \log _2\left({\rm{I}}_{p, \mathrm{DFRC}}^{\mathrm{Eve}} + \sigma_p^2 \right) \\
 & - \underbrace{\sum_{\ell}\log _2\left({\rm{I}}_{\ell}^{\mathrm{DL}} +\sigma_{\ell}^2 \right)}_{\phi_1} -\underbrace{\sum_{p} \log _2\left({\rm{S}}_{p, \mathrm{DFRC}}^{\mathrm{Eve}} + {\rm{I}}_{p, \mathrm{DFRC}}^{\mathrm{Eve}} + \sigma_p^2\right)}_{\phi_2}.
 \end{aligned}
\end{equation*}	
\useshortskip 
In this form, the \ac{DL} \ac{SSR} is not concave due to convexity of $-\phi_1$ and $-\phi_2$ terms. Thus, we intent to "convexify" the problem by introducing the following lemmas.\\
\noindent\textit{Lemma 1:} The term $\phi_1$ can be approximated to an affine function by applying Taylor expansion around $(\widetilde{\pmb{p}}, \{\widetilde{\pmb{V}}_l\}_{l=1}^L, \widetilde{\pmb{W}})$
\useshortskip
\begin{equation}
\small
\label{eq:phi1_final}
	\begin{aligned}
	\small
		\phi_1 & \approx \frac{1}{\ln 2}\sum\nolimits_{\ell}\Bigg\{\ln\left|\widetilde{{\rm{I}}}_{\ell}^{\mathrm{DL}}+\sigma_{\ell}^2\right| \\
        &+ \dfrac{\sum_k \vert q_{k,\ell} \vert^2 {\Delta}_{p_k} + \operatorname{tr}\left(\pmb{H}_{r,\ell} \left[\sum_{s \neq \ell}\pmb{\Delta}_{\pmb{V}_s} + \pmb{\Delta}_{\pmb{W}}\right]\right)}{\widetilde{{\rm{I}}}_{\ell}^{\mathrm{DL}}+\sigma_{\ell}^2}\Bigg\},
	\end{aligned}
\end{equation}
where $\small
\Delta _{p_k} = p_k - \widetilde{p}_k$, $\Delta _{\pmb{V}_s} = \pmb{V}_s - \widetilde{\pmb{V}}_s$, $\Delta _{\pmb{W}} = \pmb{W} - \widetilde{\pmb{W}}$, and $\small
\widetilde{\rm{I}}_{\ell}^{\mathrm{DL}} = \sum_{\substack{k }} \widetilde{p}_k \vert q_{k,\ell} \vert^2 +\sum_{\substack{ \ell' \neq \ell}} \trace \left( \widetilde{\pmb{V}}_{\ell'}\pmb{H}_{r,\ell} \right) +\trace \left(\widetilde{\pmb{W}}\pmb{H}_{r,\ell} \right)$.
\textbf{Proof}: Omitted due to lack of space.\\
Next, the function $\phi_2$ can also be approximated in a similar manner as $\phi_1$, which is detailed in the following lemma.
\noindent \textit{Lemma 2:} The term $\phi_2$ can be approximated to an affine function by applying Taylor expansion around $(\widetilde{\pmb{p}}, \{\widetilde{\pmb{V}}_l\}_{l=1}^L, \widetilde{\pmb{W}})$:


 \begin{equation*}
 \small
 \label{eq:DL-phi2}
	\begin{aligned}
	\small
		 \phi_2 &\approx \frac{1}{\ln 2}\sum\nolimits_{p}\Bigg\{\ln\vert \widetilde{{\rm{S}}}_{p, \mathrm{DFRC}}^{\mathrm{Eve}} + \widetilde{{\rm{I}}}_{p, \mathrm{DFRC}}^{\mathrm{Eve}} + \sigma_p^2\vert \\
        &+ \dfrac{\sum_k\vert g_{k,p} \vert^2{\Delta}_{p_k} +  \vert \alpha_p \vert^2\operatorname{tr}\left(\pmb{A}(\theta_p)\left[\sum_s\pmb{\Delta}_{\pmb{V}_s} + \pmb{\Delta}_{\pmb{W}}\right]\right)}{\widetilde{{\rm{S}}}_{p, \mathrm{DFRC}}^{\mathrm{Eve}} + \widetilde{{\rm{I}}}_{p, \mathrm{DFRC}}^{\mathrm{Eve}} + \sigma_p^2}\Bigg\},
	\end{aligned}
\end{equation*}
where $\small \widetilde{\rm{S}}_{p, \mathrm{DFRC}}^{\mathrm{Eve}} = \sum_\ell \trace \left(\vert \alpha_p \vert^2 \widetilde{\pmb{V}}_\ell\pmb{A}(\theta_p)\right)$ and $\small
 \widetilde{\rm{I}}_{p, \mathrm{DFRC}}^{\mathrm{Eve}} = 
	\sum_{\substack{k }} \widetilde{p}_k \vert g_{k,p} \vert^2 + 
	\vert \alpha_p \vert^2 \trace \left(\widetilde{\pmb{W}}\pmb{A}(\theta_p)\right)$.

\noindent \textbf{Proof:} The proof follows similar steps as that of \textit{Lemma 1}.
Finally, the complete approximation for the \ac{DL} \ac{SSR} is given at the bottom of the page in equation \eqref{eq:SR_DL_bottom}, which is readily verified to be concave in $({\pmb{p}}, \{{\pmb{V}}_l\}_{l=1}^L, {\pmb{W}})$.
\useshortskip
 \begin{figure*}[b]
 \footnotesize
 \begin{equation}
	\label{eq:SR_DL_bottom}
 \begin{aligned}
 \wideparen{\SR}_{\mathrm{DL}} &= 
 \frac{1}{\ln2}\Bigg\{\sum\nolimits_{\ell}\Bigg\{\ln\left({\rm{S}}_{\ell}^{\mathrm{DL}} + {\rm{I}}_{\ell}^{\mathrm{DL}} + \sigma_{\ell}^2 \right) - \ln\left|\widetilde{{\rm{I}}}_{\ell}^{\mathrm{DL}}+\sigma_{\ell}^2\right| - \dfrac{ \sum_k \vert q_{k,\ell} \vert^2 {\Delta}_{p_k} + \operatorname{tr}\left(\pmb{H}_{r,\ell} \left[\sum_{s \neq \ell}\pmb{\Delta}_{\pmb{V}_s} + \pmb{\Delta}_{\pmb{W}}\right]\right)}{\widetilde{{\rm{I}}}_{\ell}^{\mathrm{DL}}+\sigma_{\ell}^2}\Bigg\} \\
 & + \sum\nolimits_{p} \left.\left\{\ln\left({\rm{I}}_{p, \mathrm{DFRC}}^{\mathrm{Eve}} + \sigma_p^2 \right) - \ln\vert \widetilde{{\rm{S}}}_{p, \mathrm{DFRC}}^{\mathrm{Eve}} + \widetilde{{\rm{I}}}_{p, \mathrm{DFRC}}^{\mathrm{Eve}} + \sigma_p^2\vert - \dfrac{\sum_k\vert g_{k,p} \vert^2{\Delta}_{p_k} + \vert \alpha_p \vert^2\operatorname{tr}\left(\pmb{A}(\theta_p)\left[\sum_s\pmb{\Delta}_{\pmb{V}_s} + \pmb{\Delta}_{\pmb{W}}\right]\right)}{\widetilde{{\rm{S}}}_{p, \mathrm{DFRC}}^{\mathrm{Eve}} + \widetilde{{\rm{I}}}_{p, \mathrm{DFRC}}^{\mathrm{Eve}} + \sigma_p^2}\right\}\right\}.
 \end{aligned}
\end{equation}	
\end{figure*}
We now proceed to the \ac{UL} \ac{SSR} given in equation \eqref{eq:sum-secrecy-rate-UL}.
Note that $\Gamma_{k}^{\mathrm{UL}}$ forms a generalized Rayleigh quotient in $\pmb{u}_k$ as given in its form in \eqref{eq:SINR_UL1}. 
Also note that $\Gamma_{k}^{\mathrm{UL}}$ is the only quantity that depends on $\pmb{u}_k$. 
Based on this observation, the maximum of $\Gamma_{k}^{\mathrm{UL}}$ is attained by setting $\widehat{\pmb{u}}_k 
	=
	\pmb{\Phi}_k^{-1}(\pmb{Q},\pmb{p})
	 \pmb{h}_{t,k}, \forall k$,
and applying it in \eqref{eq:SINR_UL1}, we get $\widehat{\Gamma}_{k}^{\mathrm{UL}}
	= 
	 p_k
	 \pmb{h}^H_{t,k}
	\pmb{\Phi}_k^{-1}(\pmb{Q},\pmb{p}) 
	 \pmb{h}_{t,k}$.
Now that the \ac{UL} beamforming vectors have maximized the \ac{UL} \acp{SINR}, the maximization of the \ac{UL} \ac{SSR} can be performed based on the remaining variables, namely $({\pmb{p}}, \{{\pmb{V}}_l\}_{l=1}^L, {\pmb{W}})$, as the \ac{UL} \ac{SSR} at this point is
\useshortskip 
\begin{equation}
\label{eq:sum-secrecy-rate-UL}
\small
\widehat{\SR}_{\mathrm{UL}}
	=\sum\nolimits_{k}
	\log_2(1+\widehat{\Gamma}_{k}^{\mathrm{UL}})
	-
	\sum\nolimits_{p,k}
	\log_2(1+\Gamma_{p,k}^{\mathrm{Eve}}),
\end{equation}
\useshortskip 
which is the \ac{UL} \ac{SSR} after \ac{UL} beamforming at the \ac{DFRC} \ac{BS}. On one hand, we have less variables to maximize the \ac{UL} \ac{SSR}, but on the other hand, this rate is still highly non-convex. To "convexify" it, we first lower-bound the first term as 
\useshortskip
\begin{equation*}
\small
	\sum\nolimits_{k}
	\log_2(1+\varpi_k)
	\leq 
	\sum\nolimits_{k}
	\log_2(1+\widehat{\Gamma}_{k}^{\mathrm{UL}}),
\end{equation*} 
where we introduced $K$ slack variables satisfying $0 \leq \varpi_k \leq \widehat{\Gamma}_{k}^{\mathrm{UL}}$ for all $k$.
Next, a Taylor series expansion around $\widetilde{\boldsymbol{\Phi}}_k$ can be conducted, hence giving us a more suitable bound
\begin{equation*}
\small
	\varpi_k/p_k \leq \pmb{h}_{t,k}^H\widetilde{\boldsymbol{\Phi}}_k^{-1} \pmb{h}_{t,k}-
\pmb{h}_{t,k}^H\widetilde{\boldsymbol{\Phi}}_k^{-1}\left(\boldsymbol{\Phi}_k-\widetilde{\boldsymbol{\Phi}}_k\right)\widetilde{\boldsymbol{\Phi}}_k^{-1} \pmb{h}_{t,k}, \forall k.
\end{equation*}
However, this constraint is still non-convex due to its fractional nature, i.e. $\frac{\varpi_k}{p_k} $ is not convex in its underlying variables. Thus, we apply the following transformation by injecting slack variables $\vartheta_1 \ldots \vartheta_K$ appearing \textcolor{black}{in a quadratic structure}, i.e.
\useshortskip
\begin{equation}
\small
    \label{eq:P121}
    \tfrac{\vartheta_k^2}{p_k} \leq \pmb{h}_{t,k}^H\widetilde{\boldsymbol{\Phi}}_k^{-1} \pmb{h}_{t,k}- \pmb{h}_{t,k}^H\widetilde{\boldsymbol{\Phi}}_k^{-1}\left(\boldsymbol{\Phi}_k-\widetilde{\boldsymbol{\Phi}}_k\right)\widetilde{\boldsymbol{\Phi}}_k^{-1} \pmb{h}_{t,k}, \forall k,
\end{equation}
where the condition $\varpi_k \leq \vartheta_k^2, \forall k$ must be satisfied. 
The constraint in \eqref{eq:P121} is easily verified to be convex, but its necessary condition is not, though can be remedied through a first-order approximation around $\widetilde{\vartheta}_1 \ldots \widetilde{\vartheta}_K$ as follows
\useshortskip
\begin{equation}
\label{eq:varpiconst}
\small
\varpi_k \leq \widetilde{\vartheta}_k^2+2 \widetilde{\vartheta}_k\left(\vartheta_k-\widetilde{\vartheta}_k\right), \forall k.
\end{equation}
This is equivalent to lower bounding the \ac{UL} \ac{SSR} under \eqref{eq:P121} and \eqref{eq:varpiconst}, namely
\begin{equation}
\label{eq:sum-secrecy-rate-UL-loweer-bound}
\small
\sum\nolimits_{k}
\log_2(1+\varpi_k)
-
\sum\nolimits_{p,k}
\log_2(1+\Gamma_{p,k}^{\mathrm{Eve}})
\leq 
\widehat{\SR}_{\mathrm{UL}}.
\end{equation}
Next, we proceed in convexifying the second term appearing in the lower bound of \eqref{eq:sum-secrecy-rate-UL-loweer-bound}. Indeed, we have that
\useshortskip
\begin{equation}
\label{eq:sum-secrecy-rate-UL-loweer-bound-alternative}
\small
\begin{split}
\sum\nolimits_{k}
	\log_2(1+\varpi_k)
	&-\overbrace{\sum\nolimits_{p,k}
	\log_2({\rm{S}}_{p, k}^{\mathrm{Eve}} + {\rm{I}}_{p, k}^{\mathrm{Eve}} + \sigma_p^2)}^{K\phi_3}\\
    &+
    \sum\nolimits_{p,k}
	\log_2({\rm{I}}_{p, k}^{\mathrm{Eve}} + \sigma_p^2)
\leq 
\widehat{\SR}_{\mathrm{UL}},
\end{split}
\end{equation}
where $\small {\rm{I}}_{p, k}^{\mathrm{Eve}} = \sum_{\substack{k'\neq k}}^K p_{k'} \vert g_{k',p} \vert^2 + \vert \alpha_p \vert^2 \pmb{a}_{N_T}^H(\theta_p)\pmb{Q}\pmb{a}_{N_T}(\theta_p)$ contains the interference the $\small
 p^{th}$ eavesdropper sees and $\small
 {\rm{S}}_{p, k}^{\mathrm{Eve}} = p_k \vert g_{k,p} \vert^2$ highlights the useful part of the received at the $p^{\rm{th}}$ eavesdropper.
Define $\small
 \Psi _p^{\mathrm{Eve}} = \sum\nolimits_k {\rm{S}}_{p, k}^{\mathrm{Eve}} +\sum\nolimits_k  {\rm{I}}_{p, k}^{\mathrm{Eve}}$.
Next, due to convexity of the $\small
 -\phi_3$ term, the \ac{UL} \ac{SSR} is not necessarily concave. Thus, we apply the following lemma.

\noindent \textit{Lemma 3:} The term $\small
 \phi_3$ can be approximated to an affine function by applying Taylor expansion around $\small
 (\widetilde{\pmb{p}}, \{\widetilde{\pmb{V}}_l\}_{l=1}^L, \widetilde{\pmb{W}})$

\useshortskip
\begin{equation*}
 \label{eq:UL-phi3}
	\begin{aligned}
	\small
		 \phi_3 & \approx \frac{1}{\ln 2}\sum_{p}\Big\{\ln\vert \widetilde{\Psi} _p^{\mathrm{Eve}} + \sigma_p^2\vert \\
        &+ \dfrac{\sum_k\vert g_{k,p} \vert^2{\Delta}_{p_k} +  \vert \alpha_p \vert^2\operatorname{tr}\left(\pmb{A}(\theta_p)\left[\sum_s\pmb{\Delta}_{\pmb{V}_s} + \pmb{\Delta}_{\pmb{W}}\right]\right)}{\widetilde{\Psi} _p^{\mathrm{Eve}} + \sigma_p^2}\Big\},
	\end{aligned}
\end{equation*}
where $\small
 \widetilde{\rm{I}}_{p, k}^{\mathrm{Eve}} = \sum_{\substack{k'\neq k}}^K \widetilde{p}_{k'} \vert g_{k',p} \vert^2 + \vert \alpha_p \vert^2 \pmb{a}_{N_T}^H(\theta_p)\widetilde{\pmb{Q}}\pmb{a}_{N_T}(\theta_p)$, $\small
\widetilde{\Psi} _p^{\mathrm{Eve}} = \sum\nolimits_k \widetilde{\rm{S}}_{p, k}^{\mathrm{Eve}} +\sum\nolimits_k  \widetilde{\rm{I}}_{p, k}^{\mathrm{Eve}}$, $\small
\widetilde{\rm{S}}_{p, k}^{\mathrm{Eve}} = \widetilde{p}_k \vert g_{k,p} \vert^2$,
and  $\small
\widetilde{\pmb{Q}} = \sum_\ell \widetilde{\pmb{V}}_\ell + \widetilde{\pmb{W}}$.\\
\textbf{Proof:} The proof follows similar steps as that of \textit{Lemma 1}.

Thus, the approximated lower bound concave \ac{UL} \ac{SSR} is
\useshortskip
\begin{equation}
\label{eq:LB-UL-sum-secrecy-rate}
\begin{aligned}
\small
	&\wideparen{\SR}_{\mathrm{UL}}^{\mathrm{LB}}
	=\sum\nolimits_{k} \log_2(1+\varpi_k) +\sum\nolimits_{p,k} \log_2({\rm{I}}_{p, k}^{\mathrm{Eve}} + \sigma_p^2) \\
    &- \frac{K}{\ln 2}\sum\nolimits_{p}\Big\{\ln\vert \widetilde{\Psi} _p^{\mathrm{Eve}} + \sigma_p^2\vert \\
    &+ \dfrac{\sum_k\vert g_{k,p} \vert^2{\Delta}_{p_k} +  \vert \alpha_p \vert^2\operatorname{tr}\left(\pmb{A}(\theta_p)\left[\sum_s\pmb{\Delta}_{\pmb{V}_s} + \pmb{\Delta}_{\pmb{W}}\right]\right)}{\widetilde{\Psi} _p^{\mathrm{Eve}} + \sigma_p^2}\Big\}.
\end{aligned}
\end{equation}
Now that the lower-bound of the \ac{UL} \ac{SSR} ($\wideparen{\SR}_{\mathrm{UL}}^{\mathrm{LB}}$) is concave, we can formulate an efficient convex optimization problem that can be solved in the neighborhood of $(\widetilde{\pmb{p}}, \{\widetilde{\pmb{V}}_l\}_{l=1}^L, \widetilde{\pmb{W}})$. 
Indeed, combining the constraints in \eqref{eq:P121} and \eqref{eq:varpiconst} with problem $(\mathcal{P}_{1.1})$ in \eqref{eq:problem12}, and utilizing the concave \ac{DL} and \ac{UL} \ac{SSR}s in \eqref{eq:SR_DL_bottom} and \eqref{eq:LB-UL-sum-secrecy-rate}, we arrive at the following convex optimization problem
\useshortskip
\begin{subequations}
\label{eq:problem12} 
    \begin{eqnarray}
    \small
        (\mathcal{P}_{1.2}) & \underset{\lbrace \substack{\{\pmb{V}_\ell\}, \pmb{W},\pmb{p},\\ \{\varpi _k,\vartheta _k\}} \rbrace}{\operatorname{maximize}}& \wideparen{\SR}_{\mathrm{DL}} + \wideparen{\SR}_{\mathrm{UL}}^{\rm{LB}} \\
        &\operatorname{s.t.} & \hspace{-0.5cm}\tfrac{\operatorname{tr}\left(\left(\sum_{\ell} \boldsymbol{V}_{\ell}+\boldsymbol{W}\right) \boldsymbol{A}_s\right)}{\operatorname{tr}\left(\left(\sum_{\ell} \boldsymbol{V}_{\ell}+\boldsymbol{W}\right) \boldsymbol{A}_m\right)} \leq \operatorname{ISMR}_{\max} \label{eq:ismr-constriant}\\ 
        && \hspace{-1.5cm} \varpi _k \geq 0, \vartheta _k \geq 0, \quad \forall k\\
        && \hspace{-1.5cm} \eqref{eq:P102}, \eqref{eq:P103}, \eqref{eq:P104}, \eqref{eq:P105}, 	\eqref{eq:P111}, \eqref{eq:P112}, \eqref{eq:P121}, \eqref{eq:varpiconst}.
    \end{eqnarray}
\end{subequations}

Finally, the problem in $\small
(\mathcal{P}_{1.2})$ is convex and can be solved readily using commercially available convex optimization solvers.
Note that \eqref{eq:ismr-constriant} can be written in linear form as $\small
\operatorname{tr}\left(\left(\sum_{\ell} \boldsymbol{V}_{\ell}+\boldsymbol{W}\right) \left( \boldsymbol{A}_s - \operatorname{ISMR}_{\max} \boldsymbol{A}_m \right)\right) \leq 0$.
 To this end,
\textbf{Algorithm \ref{alg:alg1}} summarizes \ac{IJTB}, where convergence is attained by monotonically improving the cost.

									  
\begin{algorithm}[] 
 \footnotesize
\caption{IJTB for FD-ISAC UL/DL SSR maximization}\label{alg:alg1}
\begin{algorithmic}
\STATE 
\STATE {\textbf{Input}: $\lbrace \pmb{h}_{r,\ell}, \sigma_\ell \rbrace_{\ell = 1}^L$, $\lbrace \pmb{h}_{t,k}  \rbrace_{k=1}^K$, $ \lbrace \gamma_p,\theta_p , \sigma_p, \alpha_p \rbrace_{p=1}^P$, $\lbrace g_{k,p} \rbrace_{p,k=1}^{P,K}$, $\ISMR_{\mathrm{max}}$, $\beta$, $\pmb{H}_{\mathrm{SI}}$, $\pmb{A}_s$, $\pmb{A}_m$. } 
\STATE \textbf{Initialize}:
\STATE \hspace{0.5cm}  Set $m=0$,  $\pmb{V}_\ell^{(0)}= \pmb{0}$, $\pmb{W}^{(0)} = \pmb{0}$, $\pmb{p}^{(0)} =  \pmb{0}$.
\STATE \textbf{while} $ m < M_{iter}$
\STATE \hspace{0.5cm} Update $\widetilde{\pmb{Q}}$ as $\widetilde{\pmb{Q}} =\sum\nolimits_{\ell=1}^L \widetilde{\pmb{V}_\ell} + \widetilde{\pmb{W}}$.
\STATE \hspace{0.5cm} \textbf{for} $k = 1 \ldots K$
\STATE \hspace{1.0cm} Update $\pmb{\Phi}_k^{(m)}$ via \eqref{eq:phi_k_eqn} given $\widetilde{\pmb{Q}}$ and $\widetilde{\pmb{p}}$.
\STATE \hspace{1.0cm} Update $\pmb{u}_k^{(m+1)}$ as $\pmb{u}_k^{(m+1)} 
	=
	\left(\pmb{\Phi}_k^{(m)}\right)^{-1} 
	 \pmb{h}_{t,k}$.
\STATE \hspace{1.0cm} Normalize $\pmb{u}_k^{(m+1)}$ as $\pmb{u}_k^{(m+1)} \leftarrow \pmb{u}_k^{(m+1)} / \left|\pmb{u}_k^{(m+1)}\right|
$.
\STATE \hspace{0.5cm} Solve $(\mathcal{P}_{1.2}^{(m)})$ in \eqref{eq:problem12} to get $\pmb{V}_\ell^{(m+1)},\pmb{W}^{(m+1)}, \pmb{p}^{(m+1)}$.
  
\STATE \textbf{set} $\widetilde{\pmb{W}} = \pmb{W}^{(m)}$, $\widetilde{\pmb{p}} = \pmb{p}^{(m)}$, $\widetilde{\pmb{V}}_\ell = {\pmb{V}}_\ell^{(m)}$, $\widetilde{\vartheta}_k = {\vartheta}_k^{(m)} $, $\forall \ell, \forall k$.

\STATE \textbf{return}  $\lbrace \pmb{u}_k  \rbrace_{k=1}^K$, $\pmb{W}$, $\pmb{p}$, $\pmb{V}$.
\end{algorithmic}
\end{algorithm}

\section{Simulation Results}
\label{sec:simulation-results}
The \ac{DFRC} \ac{BS} is deployed at the origin of the cell with $500\meters$ radius \cite{8340227}. The residual \ac{SI} of $-110\dB$ can be achieved over \ac{RF} and digital domains of the transceiver \cite{6832464}, \cite{9887801}. 
The transmit/receive array geometries at the \ac{DFRC} \ac{BS} follow a \ac{ULA} structure with a spacing of $\lambda/2$, where $\lambda$ is the wavelength associated with the carrier frequency, which is $28\GHz$ in this case \textcolor{black}{\cite{rappaport2015millimeter}}. 
To align with realistic mmWave front-end hardware, we assume a total transmit power of $40\dBm$ \cite{8207426}. Combined with an array gain of $25\dBi$, this yields an \ac{EIRP} of $65\dBm$, which complies with standard regulatory limits.
The \ac{DFRC} transmit/receive antenna gain is $25\dBi$ \cite{8207426}.
The \ac{DL}/\ac{UL} \ac{CU}s are single-antenna users, 
where transmit and receive antenna gains are set to $17\dBi$ and $12\dBi$, respectively \cite{8207426}.
Eavesdroppers are also treated as single-antenna users, with the receive antenna gain of $12\dBi$ \cite{8207426}. Also, we utilize Rician ($k_{\rm{dB}} = 15\dB$) \textcolor{black}{\cite{8207426,rappaport2010wireless} and Rayleigh ($k_{\rm{dB}} = -\infty\dB$) \cite{9217488,rappaport2010wireless}} distributions in order to simulate various channel conditions, with $20\dB$ being the sum of all losses. The thermal noise measures $-174\dBmpHz$ \cite{8207426}, while the pathloss exponent is $2$ (free space).

We use the following schemes to analyze our \ac{IJTB} method:
\textit{(i)} \textbf{Iso AN}, where power is allocated uniformly between the \ac{AN}, the \ac{DL} beamforming and the \ac{UL} \acp{CU}; \textit{(ii)} \textbf{Iso No-AN}: The power budget is distributed equally amongst the \ac{DL} beamforming, the \ac{UL} \ac{CU}s, but no \ac{AN}; and \textit{(iii)} \textbf{Feasible}, which is a random control scheme in which the beamforming and power vectors, and \ac{AN} statistics, are randomly generated.\\
\begin{figure}[t]
	\centering
	\includegraphics[width=0.7\linewidth]{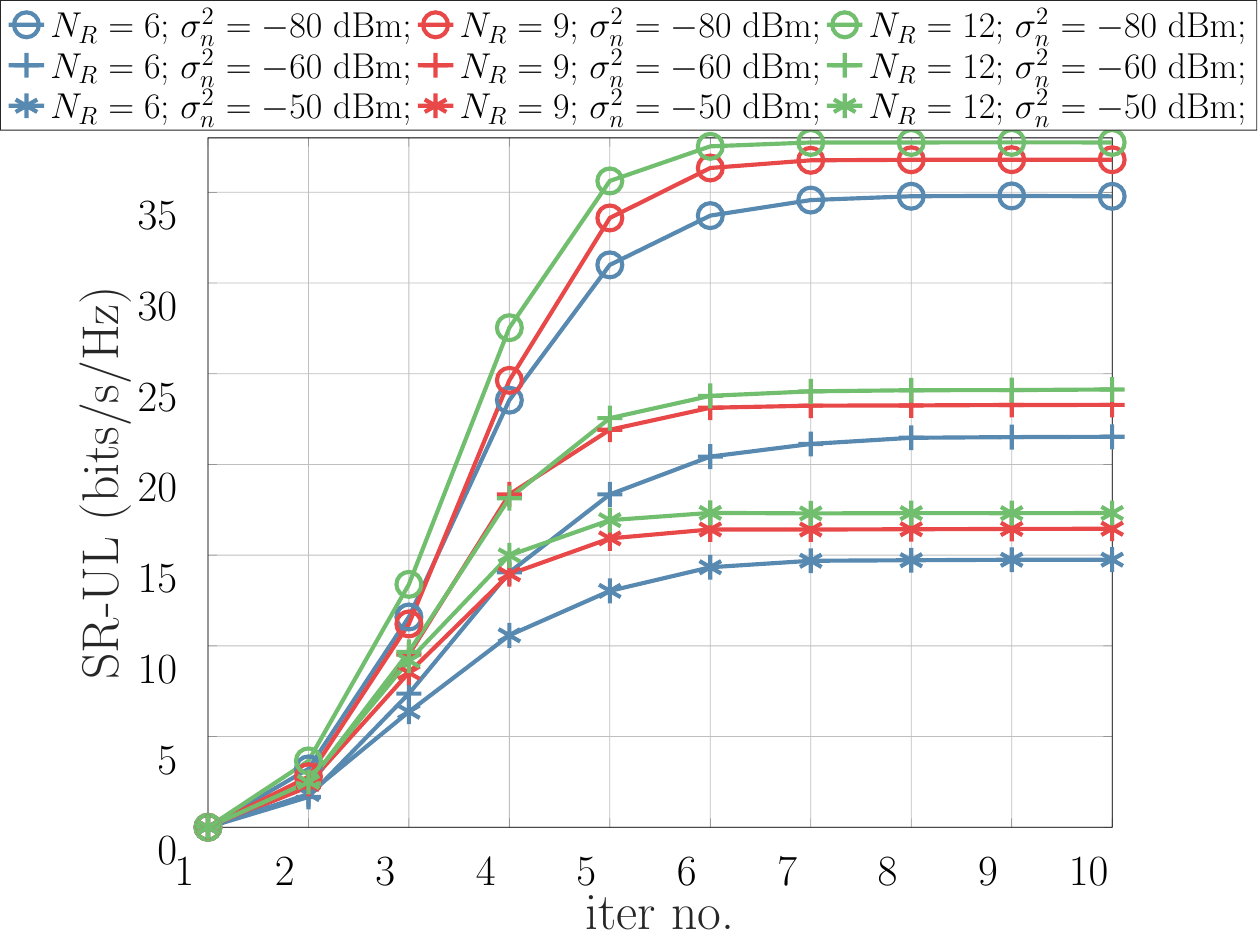}
	\caption{\ac{UL} \ac{SSR} convergence of the \ac{IJTB} method for different number of receive antennas $N_R$ and noise variance at the \ac{DFRC} \ac{BS}.}
	\label{fig:iter04}
\end{figure}
In Fig. \ref{fig:iter04}, we set $N_T = 6$, 
$N_R = 6$, 
$K = 2$, 
$L = 1$, 
$\sigma_n^2 = -70 \dB$,
$\sigma_\ell^2 = -100 \dB$,
$\sigma_p^2 = -65 \dB$,
$P_{\rm{DL}}^V = 0\dB$, 
$r_{\rm{DL}} = 20 \meters$,
$r_{\rm{UL}} = 15 \meters$, $r_{\rm{Eve}} = 17 \meters$, $P = 1$, $k_{\rm{dB}} = -\infty \dB$, and $\ISMR_{\rm{max}} = 20 \dB$.
\textcolor{black}{
The parameters in Fig.~\ref{fig:iter04} are intended for a \ac{mmWave} small cell \ac{ISAC} deployment. The carrier frequency of $28\GHz$ corresponds to the \ac{3GPP} $5$G new radio band n$257$, which was opened for $5$G for the strength of measured propagation \cite{6515173}. The transmit power of $40\dBm$ and the antenna gains of $25\dBi$ and $12\dBi$ for the \ac{DFRC} \ac{BS}, the \ac{DL}/\ac{UL} \acp{CU} and the eavesdropper follow the \ac{mmWave} link-budget recommendations in \cite{8207426}. 
The cell radius of $500\meters$ and the user/eavesdropper distances ($r_{\rm{DL}}=20\meters$, $r_{\rm{UL}}=15\meters$, $r_{\rm{Eve}}=17\meters$) reflect typical \ac{mmWave} small-cell coverage \cite{8340227}, which fall within the $500$ meter radius \cite{8340227}.
The residual \ac{SI} of $-110\dB$ corresponds to the combined \ac{RF} + analog + digital cancellation budget reported in \ac{FD} literature \cite{6832464}. 
The array sizes used in Fig.~\ref{fig:iter04}, $N_T = 6$ and $N_R \in \{6,9,12\}$, lie within the subarray dimensions of practical \ac{mmWave} hybrid-beamforming frontends and span a  range over which the \ac{DoF} benefit of additional receive antennas at the \ac{DFRC} \ac{BS} becomes visible in the \ac{UL} \ac{SSR}. 
The Rayleigh assumption ($k_{\rm{dB}} = -\infty\dB$).
The thermal noise floor of $-174\dBmpHz$ \cite{8340227,8207426} and the free-space pathloss exponent of $2$ are standard  \cite{rappaport2010wireless}. 
Finally, the receiver noise variances $\sigma_n^2 \in \{-80, -60, -50\}\dBm$ (corresponding to $-174\dBmpHz$ \cite{8207426,8340227} with appropriate bandwidth selection) swept in Fig.~\ref{fig:iter04} are deliberately chosen to span a regime so that the convergence behavior of \ac{IJTB} can be examined across distinct \ac{SNR} regimes.}

We analyze the effect of the number of \ac{DFRC} \ac{BS} receive antennas and noise variance, $\sigma_n^2$, on convergence in terms of \ac{UL} \ac{SSR}. 
For any given $\sigma_n^2$, increasing $N_R$ enhances \textcolor{black}{\ac{UL} \ac{SSR}} under a fixed power budget and \ac{ISMR} levels. 
For example, with $\sigma_n^2 = -50 \dBm$ , the \ac{IJTB} algorithm converges to an \ac{UL} \ac{SSR} of approximately $14.71 \bpspHz$, whereas it achieves $14.71 \bpspHz$ for $N_R = 6$, $16.44 \bpspHz$ for $N_R = 9$, and $17.32 \bpspHz$ for $N_R = 12$. 
The improvement arises due to extra \ac{DoF} provided by more receive antenna resources, \textcolor{black}{supporting a higher} \ac{UL} \ac{SSR} at convergence.

\textcolor{black}{In Fig.~\ref{fig:iter04}, 
increasing $N_R$ raises the converged \ac{UL} \ac{SSR} roughly with the additional receive \ac{DoF}, while lowering $\sigma_n^2$ shifts every curve upward in parallel without affecting the convergence rate. 
Increasing the number of \ac{UL} \acp{CU} $K$ or the number of eavesdroppers $P$ introduces additional interference and secrecy penalties and lowers the converged \ac{SSR}; conversely, moving the eavesdroppers farther away or the legitimate \acp{CU} closer increases the \ac{UL} \ac{SSR}. 
Tightening $\ISMR_{\max}$ forces the \ac{IJTB} solver to spend more transmit covariance adjustment on the mainlobe and reduces the converged \ac{SSR}. 
In all of these cases, the convergence behavior is qualitatively unchanged, i.e. the \ac{IJTB} curve rises monotonically and saturates within roughly less than $10$ iterations, so changing the parameters mainly shifts the plateau.}

\begin{figure}[t]
	\centering
	\includegraphics[width=0.7\linewidth]{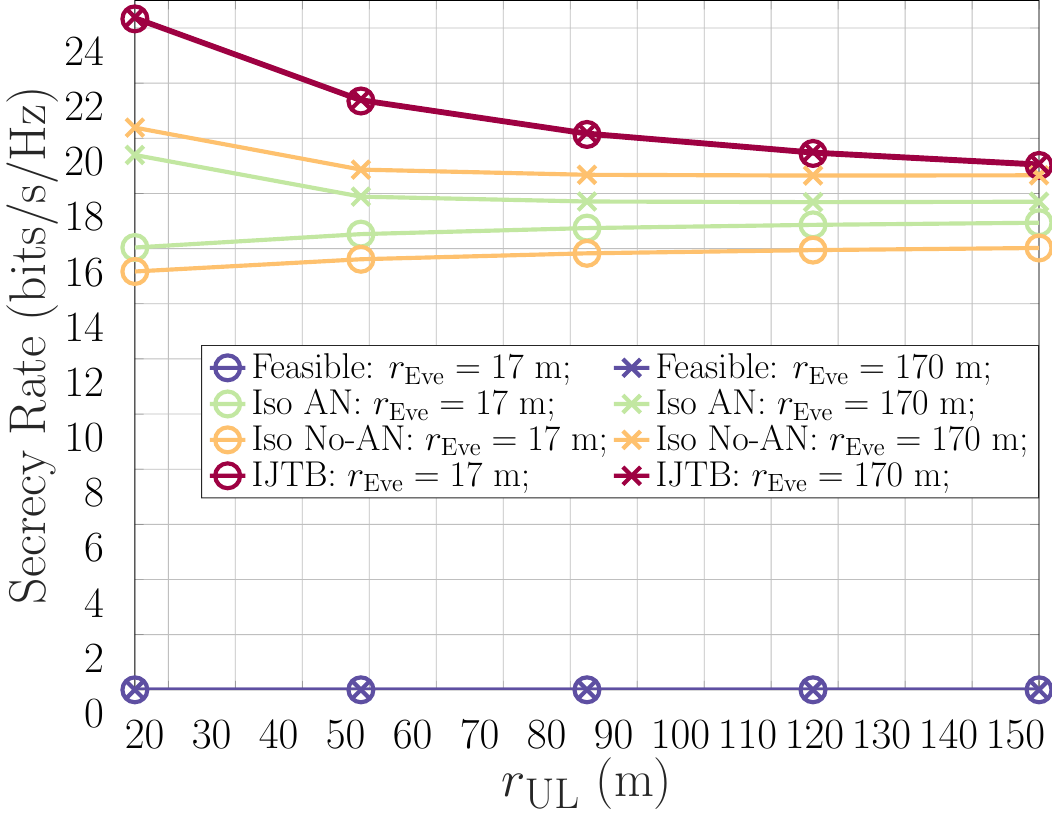}
	\caption{Overall \ac{SSR} performance for different target eavesdropper distances, i.e. $r_{\rm{Eve}}$, and for various \ac{UL} user distances.}
	\label{fig:noniter02}
\end{figure}
In Fig. \ref{fig:noniter02}, we fix the same parameters as Fig. \ref{fig:iter04}, except for the following: 
$N_R = 12$,
$K = 1$,
$P = 1$,
$k_{\rm{dB}} = -\infty \dB$,
$\sigma_\ell^2 = -100 \dB$, and
$\ISMR_{\rm{max}} = 20 \dB$. 
We further assess the \ac{IJTB} method's effectiveness \textcolor{black}{on the overall \ac{SSR}} as the \ac{UL} user distance increases, with various eavesdropper distances. We can see that \ac{SSR} decreases with \ac{UL} distance across all methods, and that far eavesdroppers generally increase \ac{SSR} due to higher path loss. Notably, with a nearby eavesdropper (e.g., $r_{\rm{Eve}} = 17$ m), \ac{AN} is more effective, yielding a $0.84$ $\bpspHz$ gain in favor of the isotropic \ac{AN} benchmark over the non-isotropic benchmarks. 
However, at greater eavesdropper distances (e.g., $r_{\rm{Eve}} = 170$ m), \ac{AN} becomes less impactful. Random solutions for power vectors, beamformers, and \ac{AN} beamforming yield an average \ac{SSR} near zero, while the \ac{IJTB} method adapts well to changing eavesdropper distances, as evidenced by similar \ac{IJTB} results across distances. It is worth noting that \textit{across all tested \ac{UL} user distances, \ac{IJTB} consistently outperforms other methods}. For instance, at $r_{\rm{UL}} = 50\meters$, the \ac{IJTB} method achieves an \ac{SSR} of approximately $21.4$ $\bpspHz$, compared to $18.86$ $\bpspHz$ for the isotropic no-\ac{AN}.\\
\begin{figure}[t]
	\centering
	\includegraphics[width=0.7\linewidth]{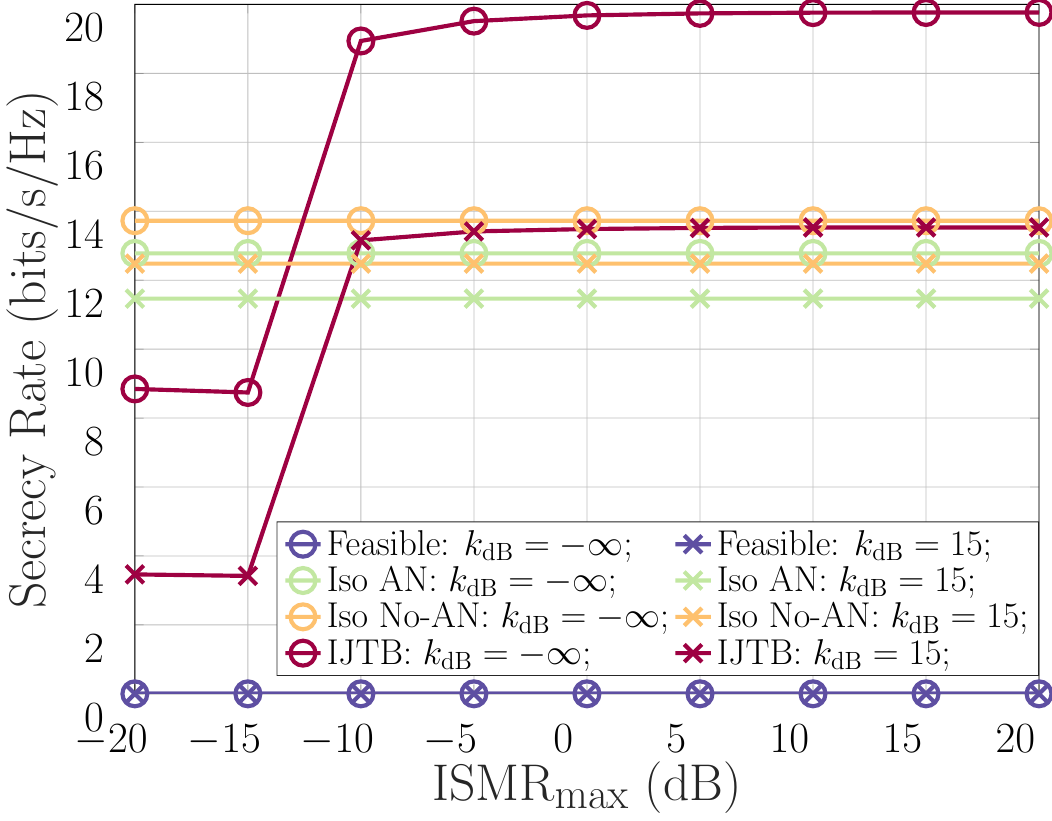}
	\caption{The \ac{SSR}-\ac{ISMR} trade-off for Rayleigh and strong \ac{LoS} channels.}
	\label{fig:noniter03}
\end{figure}
In Fig. \ref{fig:noniter03}, we fix the same parameters as Fig. \ref{fig:iter04} except for
$P = 1$,
$r_{\rm{DL}} = 200 \meters$, and
$r_{\rm{Eve}} = 175 \meters$.
We study the \ac{SSR} vs. \ac{ISMR} \ac{ISAC} trade-off for different channel conditions.
When the \ac{ISMR} constraint imposes a stringent restriction through low $\rm{ISMR}_{\rm{max}}$, the \ac{SSR} of the \ac{IJTB} method drops lower than the isotropic benchmarking methods. 
This is because the \ac{IJTB} method is dedicating an optimized chunk of power towards the eavesdropper through the mainlobes, while restricting power onto the sidelobes.
Interestingly enough, the \ac{IJTB} method is still able to achieve high \ac{SSR}s, which are, in some channel conditions, comparable with the benchmarks.
For example, in Rayleigh channel conditions ($k_{\rm{dB}} = -\infty$) and at a tight \ac{ISMR} corresponding to $\rm{ISMR}_{\rm{max}} = -20\dB$, the \ac{SSR} is about $8.84 \bpspHz$ and as the \ac{ISMR} constraint becomes less and less stringent, the \ac{SSR} of the \ac{IJTB} method gradually increases to increase the maximum possible \ac{SSR} the \ac{ISAC} system can achieve.
For example, the \ac{SSR} reaches $19.76 \bpspHz$ and $13.53 \bpspHz$ at high $\rm{ISMR}_{\rm{max}}$ for Rayleigh and strong \ac{LoS} channel conditions, respectively.
This is because the channel conditions hold significant amount of information regarding the \ac{SSR} of the \ac{IJTB} system.

\section{Conclusion}
\label{sec:conclusion}
This paper explores an \ac{FD} \ac{ISAC} system with multiple malicious targets attempting to intercept \ac{UL} and \ac{DL} communications. For \ac{FD} \ac{ISAC}, we propose an \ac{ISAC} optimization framework that maximizes \ac{UL} \& \ac{DL} \ac{SSR} while respecting sensing and power constraints, thus quantifying the highest achievable \ac{FD} secrecy rates for \ac{ISAC}. Also, we introduce the \ac{IJTB} method, tailored for \ac{FD} \ac{ISAC} secrecy rate maximization, which alternates between sub-problems to optimize beamforming and power allocations efficiently. Results confirm that the \ac{IJTB} method consistently outperforms benchmark techniques. Future work will be oriented towards privacy related \ac{ISAC} issues and secure \ac{ISAC} under calibrated \ac{AoA} sensing, using measured NYUSIM channels and hardware \ac{EIRP} limits.



\vspace{-0.2cm}
\section*{acknowledgment}
This work has been supported by Tamkeen and the Center
for Cybersecurity under the NYUAD Research Institute
Award G1104 and the Research Institute NYUAD grant CG017.

\vspace{-0.2cm}
\bibliographystyle{IEEEtran}
\bibliography{refsabr}

\vfill

\end{document}